\documentclass[epj]{svjour}

\usepackage[utf8]{inputenc}

\usepackage{graphicx}
\usepackage{soul}
\usepackage{color}

\begin{document}

\title{Electron-hole asymmetry in electrical conductivity of low-fluorinated graphene: Numerical study}
\titlerunning{Electron-hole asymmetry in electrical conductivity of low-fluorinated graphene}
\authorrunning{D.V. Kolesnikov \and V.A. Osipov}
\author{D.V. Kolesnikov\inst{1,2} \and V.A. Osipov\inst{1}}                     
\institute{ Bogoliubov Laboratory of Theoretical Physics,  Joint Institute for Nuclear Research, Dubna, Russia \and e-mail:kolesnik@theor.jinr.ru}
\date{Received: date / Revised version: date}
%
\abstract{
By using the real-space Green-Kubo formalism we study numerically the electron transport properties of low-fluorinated graphene. At low temperatures the diffuse transport regime is expected to be prevalent, and we found a pronounced electron-hole asymmetry in electrical conductivity as a result of quasi-resonant scattering on the localized states. For the finite temperatures in the variable-range hopping transport regime the interpretation of numerical results leads to the appearance of local minima and maxima of the resistance near the energies of the localized states. A comparison with the experimental measurements of the resistance in graphene samples with various fluorination degrees is discussed.
} 

\maketitle

\section{Introduction}

\quad A theoretical investigation of novel 2d materials is an important but challenging task. The first synthesized  2d material, graphene, continues to be relevant today. Possible applications of graphene in various devices for high-speed electronics\cite{hiel,hiel2} require significant modification of its electronic properties. 

Adatom adsorbtion on graphene sheet produces more stable samples and greatly influences the physical characteristics such as electrical and thermal conductivity, plasmonics, optics, etc.\cite{etc0} The recent progress in the experiments with fluorinated graphene includes synthesis of partially fluorinated samples with controlled degree of fluorination (see, e.g., reviews~\cite{okotrub,exp05}). In turn, this is what allows precise tuning of the electronic bandgap~\cite{gap1}.

Various factors can influence transport properties of a micrometer-sized chemically modified graphene sample. Among them are the presence of the electronic states localized on the adatoms, which can occur both inside the gap and in one of the bands and induce either resonant (in case of passivated graphene)~\cite{cochan} or non-resonant scattering. This is of importance in the diffuse transport regime. For fluorinated graphene, special attention should be paid to the spin-orbit interaction~\cite{fg-soc}, especially for fluorinated graphene nanoribbons~\cite{fgribbon}, where the bandgap may arise both due to fluorination, and from the geometrical quantization. In particular, the role of spin-flip process in the electrical transport is highlighted in~\cite{pribbon} where the existence of additional transport channel within the bandgap was shown.  Another factor is the localization due to multiple scattering on defects and quantum interference that gives rise to the phenomena of localization and hopping transport. Several theoretical models for the description of adatoms on graphene were formulated~\cite{midgap,ire,yako}. Due to mutual influence of the  aforementioned effects, it is necessary to use modern numerical methods for the correct description of the density of electronic states, conductivity and mobility in fluorinated graphene. Such methods should combine large size of the supercell with the Hamiltonian derived from the first-principles calculation, which takes into account the spin-orbit interaction. In this paper, we use a real-space Green-Kubo method to calculate electronic and transport properties of partally fluorinated graphene both in diffuse and hopping transport regimes. 

Our paper is organized as follows. In Section 2, the computational formalism is discussed. The tight-binding Hamiltonian for the fluorinated graphene, which includes the spin-orbit interaction is described and the real-space Green-Kubo method is introduced. In Section 3, the numerical results for the diffuse transport regime are shown and a comparison with analytic results for the density of states is discussed.  The case of finite-temperature variable-range hopping transport is considered in Section 4. Section 5 is devoted to main conclusions.
\section{Computational formalism}

We start with the Green-Kubo formula for the zero-temperature electrical conductivity
\begin{equation}
\sigma^{GK}_{\mu\nu}(E)=\frac{2\pi\hbar e^2}{\Omega} Tr[\hat{V}_\mu \delta(E-H)\hat{V}_\nu \delta(E-H)],
\end{equation}
where $\sigma^{GK}$ is the conductivity tensor, $\hbar$ is the reduced Planck constant, $e$ is the electron charge, 
$\Omega$ is the 2D volume, $\delta$ represents the delta function, $E$ is the energy, $H$ is the Hamiltonian of the system and $\hat V_\mu, \mu=x,y$ is the velocity operator in the $\mu$ direction. One can use the Fourier transformation of one delta-function $\delta (E-H) = \int_{-\infty}^\infty \exp(i(E-H)t/ \hbar) dt/{2\pi\hbar},$ so that
the conductivity is written as 
\begin{equation}
\sigma^{GK} = \frac{e^2}{\Omega}\int_{-\infty}^\infty dt Tr[e^{iHt}\hat{V} e^{-iHt}\hat{V}\delta(E-H)],
\end{equation}
where $\hat{V}$ is the velocity operator in the x direction. Using Eq.(2), one can get the Green-Kubo formula for the running electrical conductivity $\sigma^{GK}(E,t)$ and the density of states (DOS) $\rho(E)$ from the velocity autocorrelation function $C_{vv}$ as
\begin{eqnarray}
\sigma^{GK} (E,t)=e^2 \rho(E) \int_0^t C_{vv}(E,t) dt,\\
C_{vv}(E,t) = \frac{Tr[\frac{2}{\Omega} \delta(E-H)(\hat{V}(t)\hat{V} + \hat{V}\hat{V}(t))/2]}{Tr[\frac{2}{\Omega}\delta(E-H)]},\\
\rho(E) = Tr[\frac{2}{\Omega}\delta(E-H)],
\end{eqnarray} 
where $\hat{V}(t) = \hat{U}^\dagger(t)\hat{V}\hat{U}(t) = e^{iHt/\hbar}\hat{V}e^{-iHt/\hbar}$ is the velocity operator in the Heisenberg representation and $\rho(E)$ is the density of states. Furthermore, by integrating the Green-Kubo formula, one can 
find the Einstein formula for the running electrical conductivity (REC) as a derivative of the mean square displacement (MSD)  $\Delta X^2(E,t)$, which is also known as Roche-Mayou formula \cite{roma}
\begin{eqnarray}
\sigma^E(E,t) = e^2 \rho(E) \frac{1}{2}\frac{d}{dt}\Delta X^2(E,t),\label{sigmaE}\\
\Delta X^2(E,t) = \frac{Tr[\frac{2}{\Omega}\delta(E-H)(\hat{X}(t)-\hat{X})^2]}{Tr[\frac{2}{\Omega}\delta(E-H)]},
\end{eqnarray}
where $\hat{X}(t)=\hat{U}^\dagger (t)\hat{X}\hat{U}(t)$ is the x-coordinate operator in the Heisenberg representation. To estimate the conductivity from the running conductivity, we use the large time limit $\sigma(E)=\lim_{t\rightarrow \infty}\sigma^{E}(E,t)$. One can also estimate the transport regime from the running electrical conductivity by comparing the REC with the mean square displacement 
\begin{equation}
\Delta X(E,t) = \sqrt{\Delta X^2(E,t)},\label{deltax}
\end{equation}
where in the ballistic regime the REC would be constant, in the diffuse transport regime $\sigma^E\sim {\Delta X}^{-1}$ (the resistance increases linearly with the length of the sample), and in the localized regime $\sigma^E\sim \exp(-\Delta X/\zeta)$ with $\zeta$ being the localization length. Notice, however, that in real samples of finite size the intermediate regime can occur (see \cite{RadS}).

For the numerical calculation of the conductivity from the  Roche-Mayou formula (\ref{sigmaE}), the kernel polynomial method is used~\cite{KPM}. First, we normalize the Hamiltonian, energy and time as $H\rightarrow H/\Delta E, E\rightarrow E/\Delta E, t\rightarrow t \Delta E$, where $\Delta E$  is the scaling energy parameter, so that the eigenvalues of scaled Hamiltonian lies within the interval [-1,1]. By expanding the delta function in (\ref{sigmaE}) by the Chebyshev polynomials $T_n(E), n=0\ldots N_m-1$,  one can find 

\begin{eqnarray}
\rho(E)=A\sum_{n=0}^{N_m-1}g_n(2-\delta_{n0})T_n(E)C_n^{DOS},\label{rhoE}\quad\\
\rho(E)\Delta X^2(E,t)=A\sum_{n=0}^{N_m-1}g_n(2-\delta_{n0})T_n(E)C_n^{MSD}(t),\quad \\
A=\frac{2}{\pi\Omega\Delta E\sqrt{1-E^2}},\quad
\end{eqnarray}
where  $g_n=(1-n\alpha)\cos (\pi n \alpha) + \alpha \sin (\pi n \alpha)\cot (\pi\alpha)$ is the Jackson damping function, $\alpha = 1/(N_m + 1)$, $\delta_{ij}$ is the Kronecker symbol, $T_n(E)$ is the nth Chebyshev polynomial, and $C_n^{DOS}, C_n^{MSD}(t)$ are the Chebyshev moments. The moments are found by the kernel polynomial method \cite{KPM}  using $N_r$ random vectors $|\phi>$
\begin{eqnarray}
C_n^{DOS} \approx <\phi|T_n(H)|\phi>,\label{Cn}\\
C_n^{MSD} \approx <\phi|[\hat{X},\hat{U}(t)]^\dagger T_n(H)\hat{U}^\dagger [\hat{X},\hat{U}(t)]|\phi>\label{Cmsd}.
\end{eqnarray} 
By using (\ref{rhoE}), (\ref{sigmaE}) and (\ref{Cn}), (\ref{Cmsd}) the density of states and electrical conductivity are estimated. To calculate the correlators $[\hat{U}(t),X]$, the recursive algorithm is used (see \cite{KG-Gr,GPUQT}). Note that by limiting the 
delta function expansion with $N_m$ Chebyshev polynomials we achieve the finite energy resolution determined as $dE\approx\pi\Delta E/N_m$. As for the estimation of the relative error due to the finite number of random vectors, it decreases with  $N_r$ increasing as $1/\sqrt{N N_R}$, where $N$ is the dimension of the Hamiltonian. Therefore, the real-space Green-Kubo method is especially useful for large systems. 

To describe the electronic properties of fluorinated graphene, a tight-binding approach with inclusion of the phenomenological spin-orbit coupling terms is implemented \cite{fg-soc}. While the spin-orbit interaction constant in pristine and polycrystalline graphene is estimated as $\approx 10\;\mu$eV~\cite{grain14}, in the chemically modified graphene it increases to several meV. In the latter case, accounting for the spin-orbit interaction in the highly fluorinated samples leads to the new spin-flip scattering channel in the band gap~\cite{fgribbon} playing major role in the electron transport.  The tight-binding model includes the $\pi$-orbitals of carbon atoms (with annihilation operators $\hat{c}_{i,\sigma}$ for atom i and spin $\sigma$) and $p_z$-orbital of the fluorine atoms on the top of the carbon atom i with annihilation operators $\hat{F}_i$. The total Hamiltonian of the system can be obtained in the form
\begin{equation}
H = H_0 + H' + H_{SO},\label{totalH}
\end{equation}
where $H_0$ is the Hamiltonian of the pristine graphene
\begin{equation}
H_0 = t \sum_{<i,j>}\sum_\sigma (\hat{c}^\dagger_{i,\sigma}\hat{c}_{j,\sigma} + \hat{c}^\dagger_{j,\sigma}\hat{c}_{i,\sigma}),\label{H0}
\end{equation}
where t=2.6 eV is the orbital hopping parameter, the sum runs over all the nearest neighboring carbon atoms $<i,j>$ (A and B type), $H'$ describes the interaction between fluorine and carbon atoms
\begin{equation}
 H' = \epsilon_F \sum_k\sum_\sigma \hat{F}^\dagger_k \hat{F}_k + T \sum_k \sum_\sigma ( \hat{F}^\dagger_k \hat{c}_k + \hat{c}^\dagger_k \hat{F}_k).
\end{equation} 
Here k runs over all the fluorinated carbon atoms, $\epsilon_F$ is the on-site energy on the fluorine adatom, and T is the orbital hopping term between the fluorine adatom and the fluorinated carbon. The spin-orbit term $H_{SO} = \sum_k H_{SO,k}$ takes into account the spin-orbit interaction in the fluorinated graphene. For the atom k with (say) type A the spin-orbit terms include interactions between $k$ and its nearest-neighboring B-type atoms $k'$, as well as the two terms representing interactions between $k'$:
\begin{eqnarray}
H_{SO, k} = \frac{i\Lambda^B_I}{3\sqrt{3}} \sum_{<<k',k''>>} \sum_\sigma \hat{c}^\dagger_{k',\sigma}\nu_{k',k''}(\hat{s}_z)_{\sigma\sigma}\hat{c}_{k'',\sigma}+\nonumber\\
+\frac{2i\Lambda_R}{3}\sum_{k'\in k_{nn}}\sum_{\sigma\neq\sigma'} (\hat{c}^\dagger_{k,\sigma}(\hat{s} \times \vec{d}_{k k'} )_{z,\sigma\sigma'}\hat{c}_{k',\sigma}  + H.c. ) +\nonumber\\
+\frac{2i\Lambda^B_{PIA}}{3} \sum_{<<k',k''>>} \sum_{\sigma\neq\sigma'} \hat{c}^\dagger_{k',\sigma}(\hat{s} \times \vec{d}_{k', k''} )_{z,\sigma\sigma'}\hat{c}_{k'',\sigma}
\end{eqnarray}
where B-type atoms k', k'' are the nearest neighbours  ($k'\in k_{nn}$) for the A-type fluorinated atom k and 2nd neighbours for each other (which is represented by the summation $<<k',k''>>$), $\vec{d}_{ij}$ is the vector connecting sites i and j, and $\nu_{ij}=\pm 1$ if the path ikj is counterclockwise (clockwise), see \cite{fg-soc} for detail. 
We include the equal number of randomly distributed fluorinated atoms in the system with fluorine concentrations 0.5\% and 1\%. To ensure the self-consistency of the scheme, no two fluorinated carbon atoms are the nearest neighbors. The parameters for the Hamiltonian are presented in \ref{table1}. We have calculated the density of states, electrical conductivity and electron mobility of the samples.

\begin{table}[h]
\begin{center}
\begin{tabular}{llllll}
F concentration & $T$, eV & $\epsilon_F$, eV & $\Lambda^B_I$, meV & $\Lambda^B_{PIA}$, meV & $\Lambda^B_R$, meV \\ \hline
0.5\%  & 6.1 & -3.3 & 3.2 & 7.9 & 11.3 \\ \hline
1\% & 5.5 & -3.2 & 3.3 & 7.3 & 11.2 
\end{tabular}
\end{center}
\label{table1}
\end{table}

 To provide comparison with the calculations of the density of states, an analytical approximation for the change in the density of states based on the Löwdin-Schrieffer-Wolff transformation is used:
\begin{equation}
\delta\nu(E) = \frac{1}{\pi} Im [\frac{\alpha(E)}{1-\alpha(E)G_0(E)}\frac{\partial}{\partial E}G_0(E)],\label{rhoA}
\end{equation}
where $\alpha(E)=T^2/(E-\epsilon_F)$, and the Green's function per spin per atom can be approximated with the function
\begin{equation}
G_0(E) \approx \frac{E}{D^2}[ ln | \frac{E^2}{D^2-E^2} | - i\pi sgn(E)\Theta(|D-E|) ] ,
\end{equation}
where $D=t\sqrt{\sqrt{3}\pi}\approx 6$ eV (see \cite{fg-soc,kondop}). Furthermore, the electron (hole) carrier density was calculated for T=300 K from the density of states as
\begin{eqnarray}
  n_h(E_F) = \int_{-\infty}^{E_D} \rho(E)(1-f(E-E_F))dE,\label{nh}\\
   n_e(E_F) = \int_{E_D}^\infty \rho(E)f(E-E_F)dE,\label{ne}
\end{eqnarray}
where $E_D$ is the graphene Fermi energy at zero gate voltage, and $f$ is the Fermi-Dirac distribution function. The conductivity can also be expressed as a function of the carrier mobility $\mu$ and density $n_i,\; i=e,h$ as $\sigma=e n_i\mu$, so that the mobility is found as $\mu(E)=\sigma(E)/(e n_i)$.

\section{Numerical results}

We have calculated the density of electronic states, electrical conductivity and electron mobility by the numerical methods described above. For the DoS calculations we used $N_m$ = 1600 Chebyshev moments and $N_r$=32 random states, while for the conductivity and mobility calculations we used $N_m$=900 moments and $N_r$=64 random states. We have performed the time evolution for the times $t\leq$ 120 fs.
\begin{figure}[ht!]
\includegraphics[height=7cm]{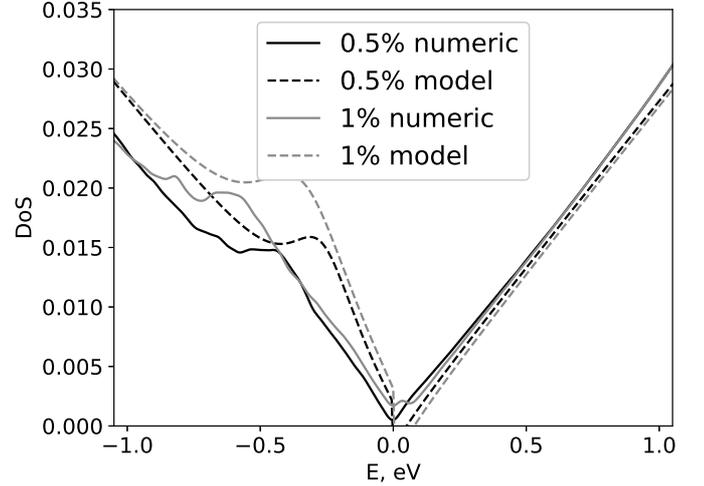}
\caption{The density of states (per atom, per eV) as a function of energy, found numerically using (\ref{rhoE}) and with analytic approximation (\ref{rhoA}).}
\label{fig1}
\end{figure}
The density of states (per eV per atom) as a function of energy is shown in Fig.\ref{fig1}. One can see the increase in DoS at E$\approx$-0.3 eV and E$\approx$-0.6 eV for 0.5\% and 1\% fluorine concentrations, respectively. For the positive energies, the DoS is close to that in pristine graphene. The peaks in the density corresponding to the localized states appear at the energies lower than those predicted by the analytical model (\ref{rhoA}) (dashed lines, see \cite{fg-soc}). In our calculations, the energy scale resolution can be approximated as 0.02 eV, which does not allow us to examine the detailed structure of the DoS at E=0. Nevertheless, one can see some finite DoS increasing with the increase of adatom concentration, which contradicts the existence of a small band gap in the analytical model. One should note that the absence of the DoS at E=0 in the analytical description is an artefact of the model approach, where the elimination of the defects within the Löwdin-Schrieffer-Wolff transformation eliminates the localized states in the band gap as well.

\begin{figure*}
\begin{center}
\resizebox{0.9\textwidth}{!}{
\includegraphics{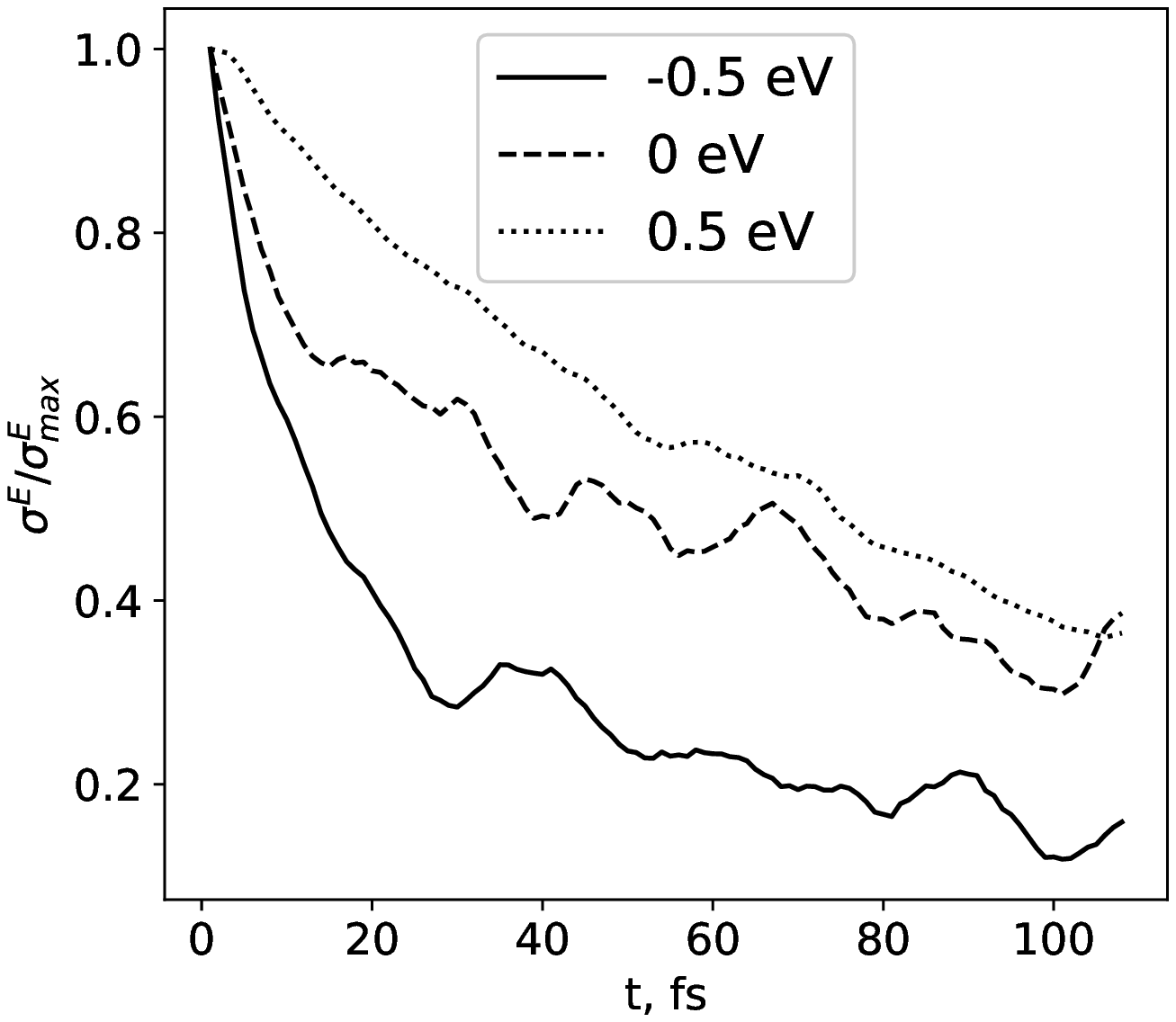} \includegraphics{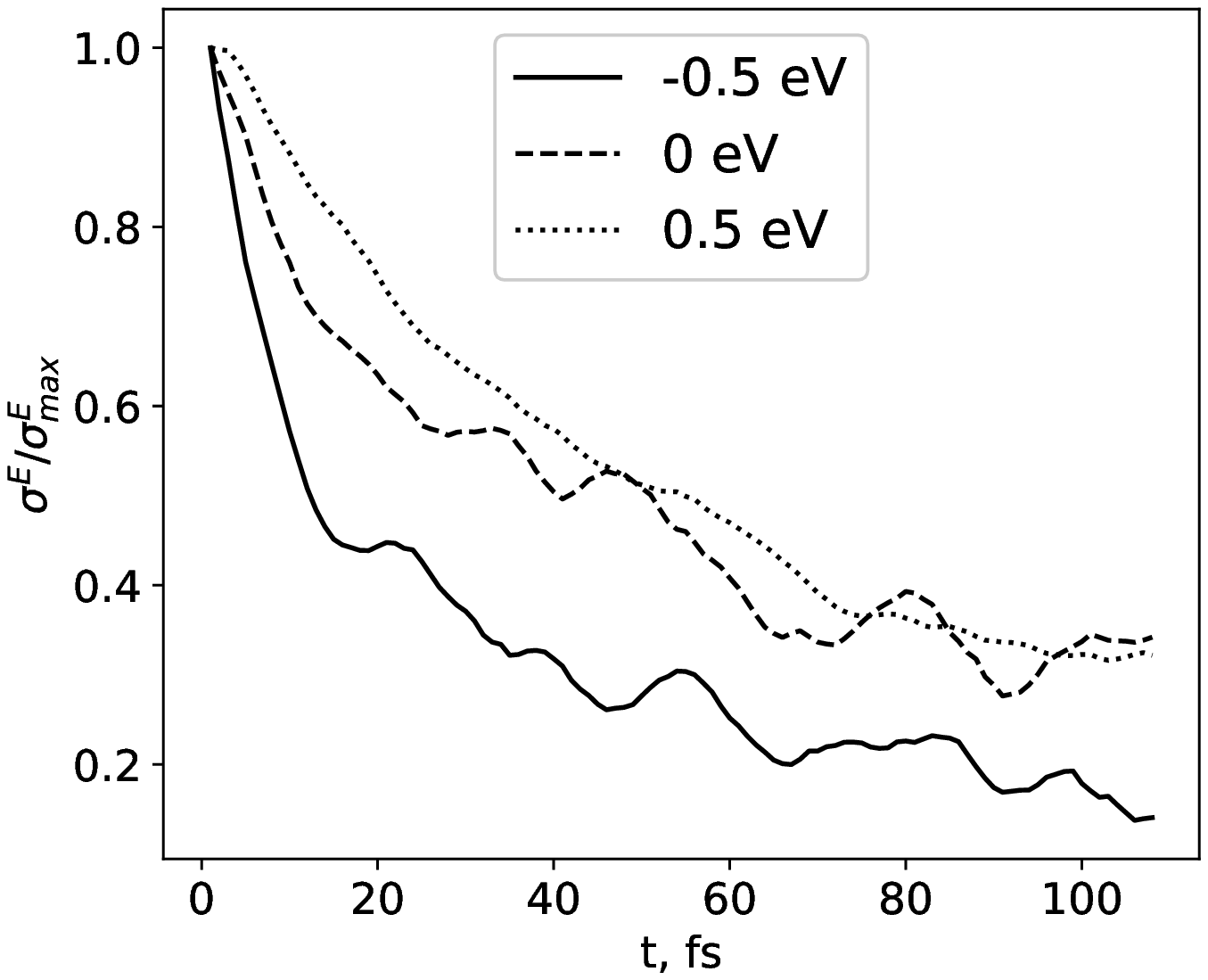} }
\end{center}
\caption{The normalized running electrical conductivity $\sigma^E(E,t)/\sigma_{max}(E)$ vs time (fs), for 0.5\% (left) and 1\% (right) fluorine concentration.}
\label{fig2}
\end{figure*}
Fig.\ref{fig2} shows the normalized value of running electrical conductivity (REC, $\sigma^E(E,t)$) as a function of simulation time (in fs) for energies $E=$-0.5, 0 and 0.5 eV. One can see the conductivity to converge at times $t>100$ fs, which is more pronounced for the positive values of energy. 

\begin{figure*}
\begin{center}
    \resizebox{0.9\textwidth}{!}{
    
\includegraphics{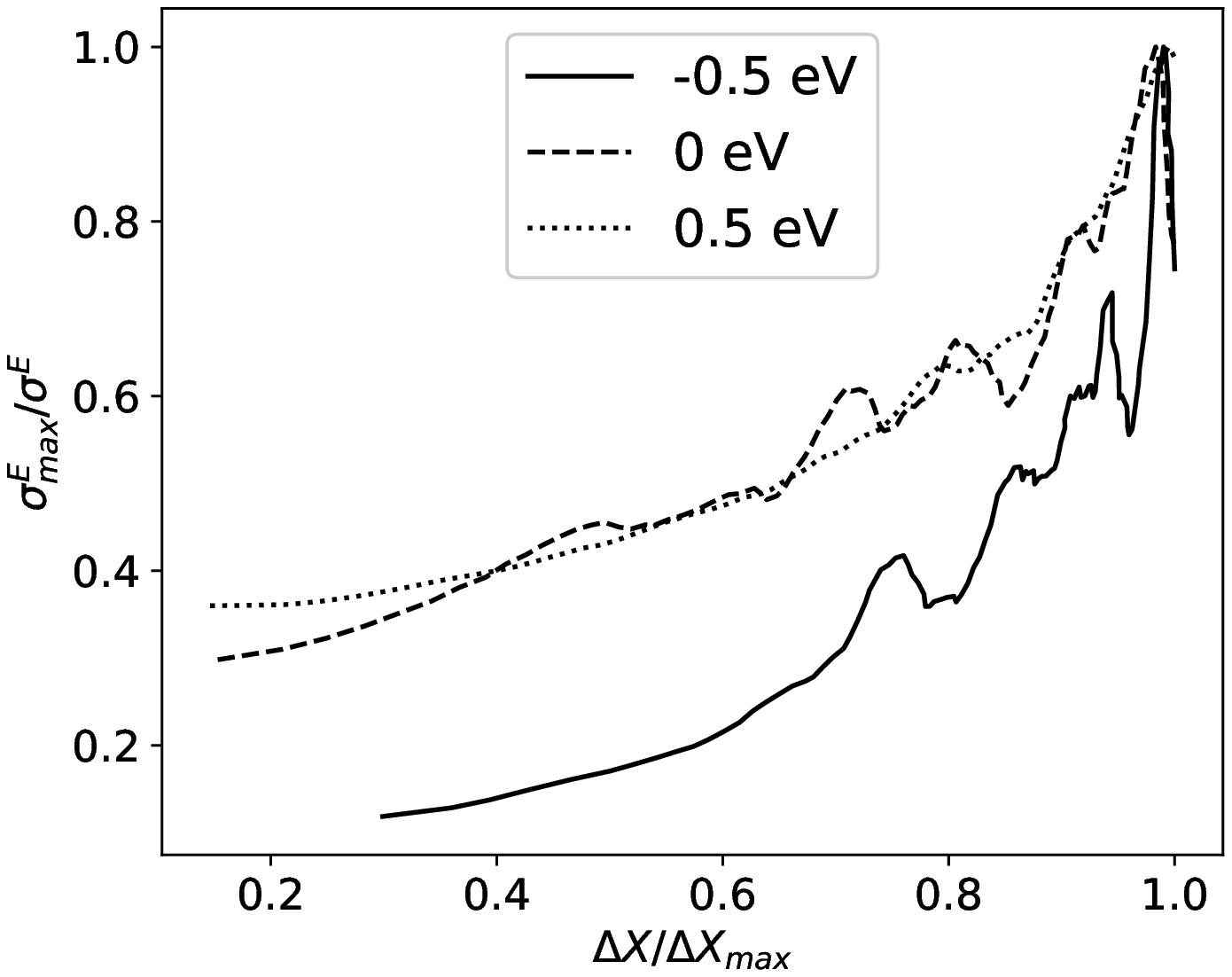} \includegraphics{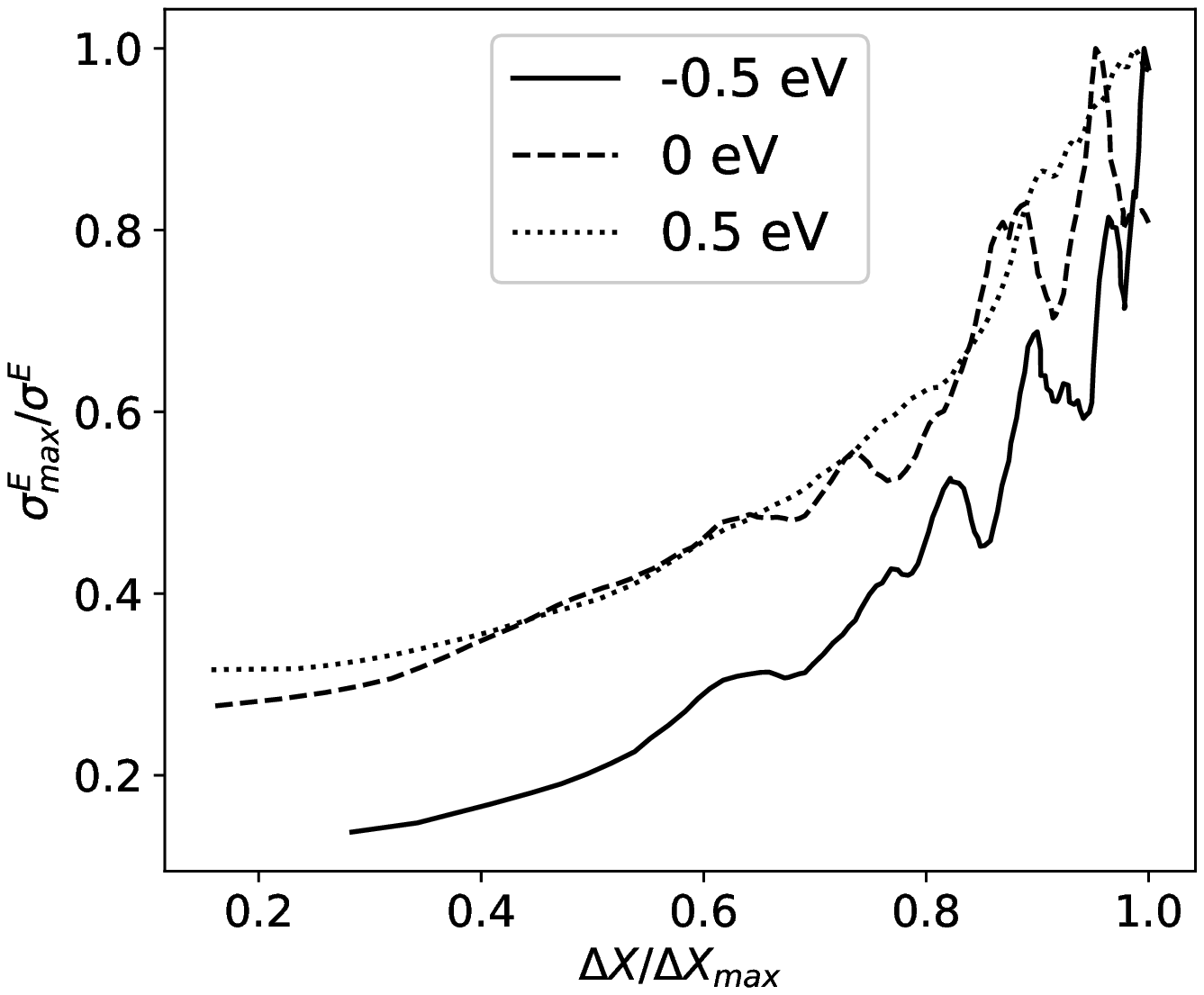} }
\end{center}

\caption{The normalized inverse running electrical conductivity $1/\sigma^E(E,t)$ vs the normalized mean square displacement $\Delta X(E,t)/\Delta X_{max}(E)$, for 0.5\% (left) and 1\% (right) fluorination.}
\label{fig3}
\end{figure*}
The normalized inverse running electrical conductivity $1/\sigma^E(E,t)$ vs the normalized mean square displacement is shown in Fig. \ref{fig3} (see (\ref{deltax})). The diffuse transport regime with the (almost) linear dependence of the inverse REC on the MSD is present for considered simulation times ($t>50$ fs) and system sizes ($\Delta X\approx 1..5\mu m$). For larger simulation times, some characteristics of the localized regime can be observed.

\begin{figure}
\includegraphics[height=7cm]{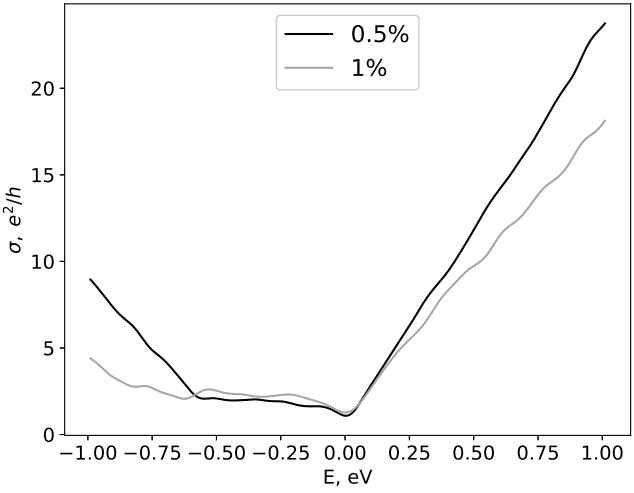} 
\caption{The electrical conductivity (in $e^2/h$) as a function of energy.}
\label{fig4}
\end{figure}
The electrical conductivity as a function of energy is shown in Fig.\ref{fig4}. As seen, the conductivity decreases with increasing concentration of adatoms and is almost linear in energy for $E>0$. For the negative energies, the conductivity is markedly reduced due to scattering on the localized states and has highly non-linear behavior. More specifically, the conductivity is significantly decreased in a wide energy interval whose center is at the peak of DoS (see Fig.\ref{fig1}) and the width  significantly exceeds the half-width of the peak. 
\begin{figure}
\includegraphics[height=7cm]{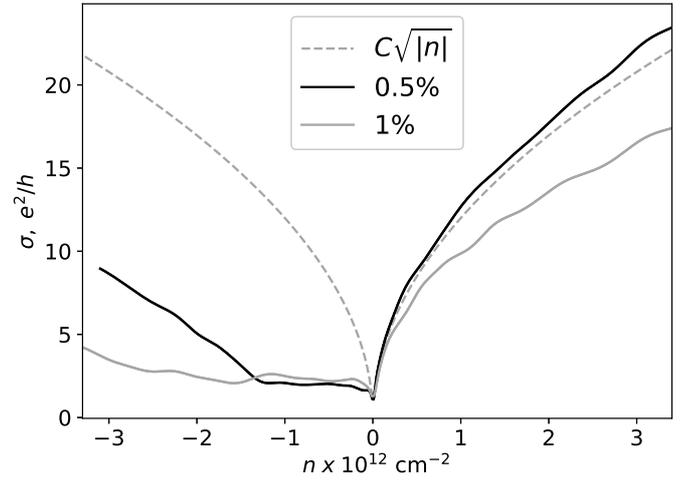} 
\caption{The electrical conductivity  as a function of carrier density, where negative (positive) density represents holes (electrons), respectively.}
\label{fig5}
\end{figure}
Fig.\ref{fig5} shows the electrical conductivity as a function of the carrier density $n$ for the holes ($n<0$) and electrons. For electrons, the square-root  behaviour of the conductivity vs the carrier  concentration is found, which is characteristic of diffuse scattering. For holes, the conductivity has a non-linear dependence on the carrier concentration due to the scattering on the localized states. 
\begin{figure*}
\begin{center}
\resizebox{0.9\textwidth}{!}{
\includegraphics[height=7cm]{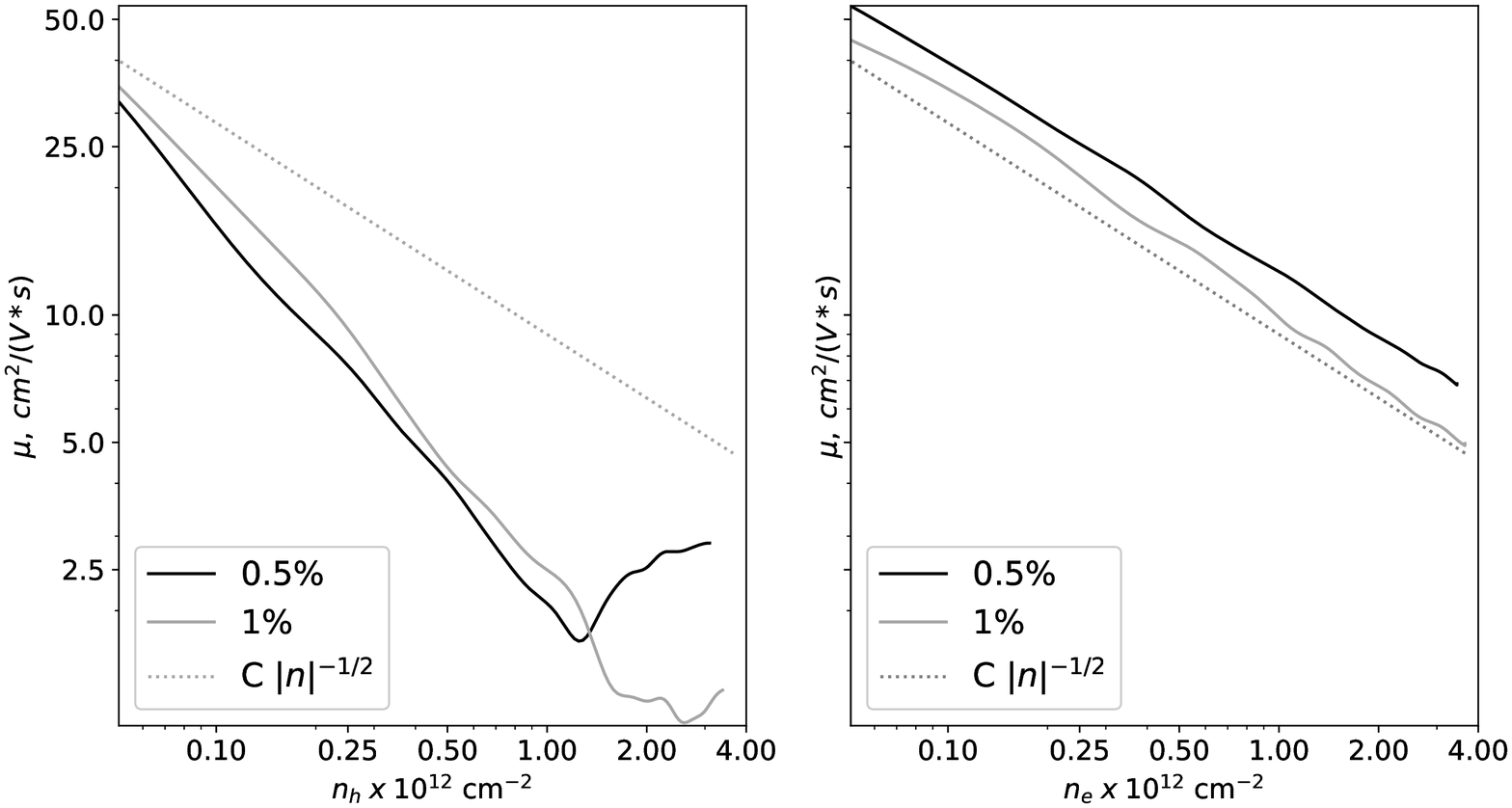} }
\end{center}
\caption{The electron mobility  as a function of carrier density for electrons (left) and holes (right).}
\label{fig6}
\end{figure*}
In Fig.\ref{fig6} the carrier mobility as a function of carrier concentration is present. One can see the clear $\mu\sim n_e^{-1/2}$ behaviour for the case of electrons, which is typical for the diffusive transport regime. The mobility of holes does not follow the diffusive behaviour, but rather follows the power-law one $n_h^\alpha$, where $ -1/2\leq \alpha\leq 0$ for low concentrations and is reduced for higher concentrations. One should also note the slight increase of the mobility with the concentration of the adatoms increasing for low hole concentrations in contrast with the case of electrons.

It is necessary to emphasize the existence of two important factors that were not included in the Green-Kubo scheme presented above. The first one is a finite size of the graphene sample. Within the real-time Green-Kubo formalism we calculate the conductivity for certain values of the time ensuring the presence of diffuse transport regime for times considered. Nevertheless, the mean square deviation may exceed the length of the sample and thus the transport in a sample of finite size can be a mixture of diffuse and localized  regimes at different energies. The second factor is the presence of temperature fluctuations that are enabling the hopping transport regime. At small temperatures, the diffusion coefficient as a function of time (and therefore, the REC) determines the  transport regime \cite{gr-qtrans}: with the elastic mean free path being much smaller than the localization length, the diffusion coefficient tends to be constant at larger times thus leading to the diffusive regime. On the other hand, when the localization length determined by the quantum interference is compatible with elastic scattering mean free path, localization occurs, which leads to the exponential decay of the diffusion coefficient. When the temperature is increasing, the hopping transport mechanism becomes prevalent, which will be discussed below.

 \section{Finite-temperature transport properties}

\begin{figure*}
\begin{center}
\resizebox{0.9\textwidth}{!}{
\includegraphics[width=16cm]{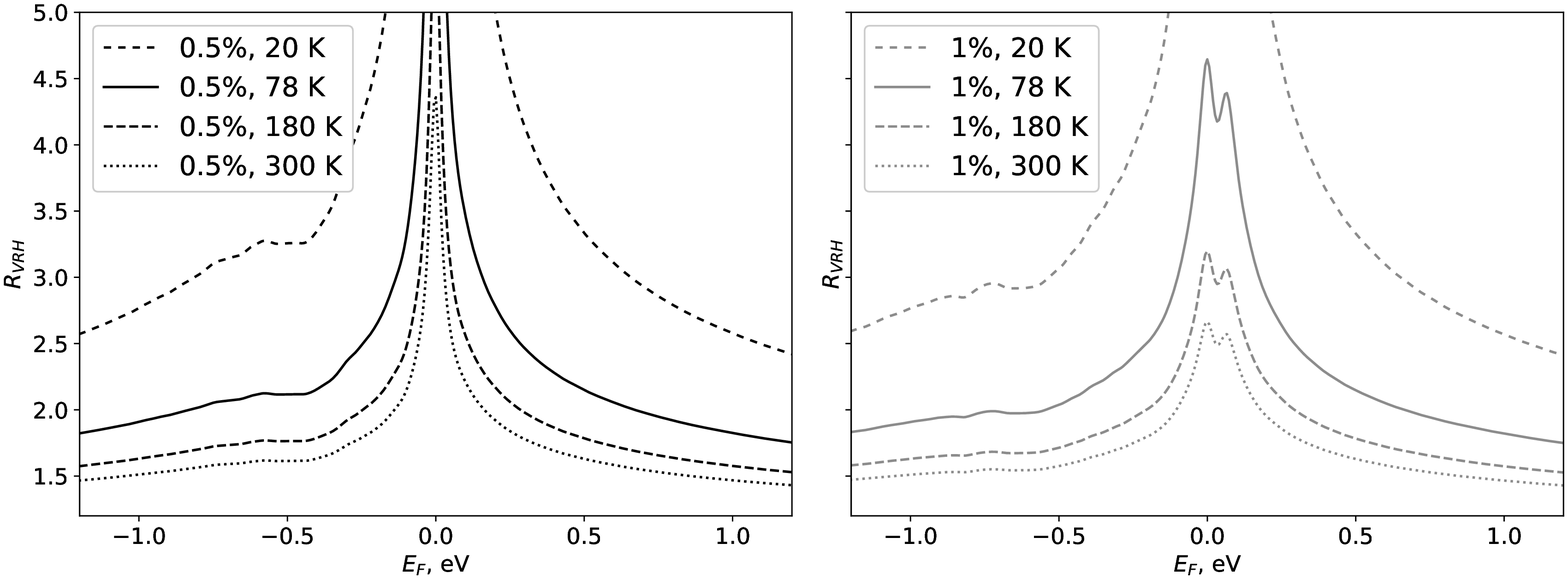} }\\
\resizebox{0.9\textwidth}{!}{
\includegraphics[height=7cm]{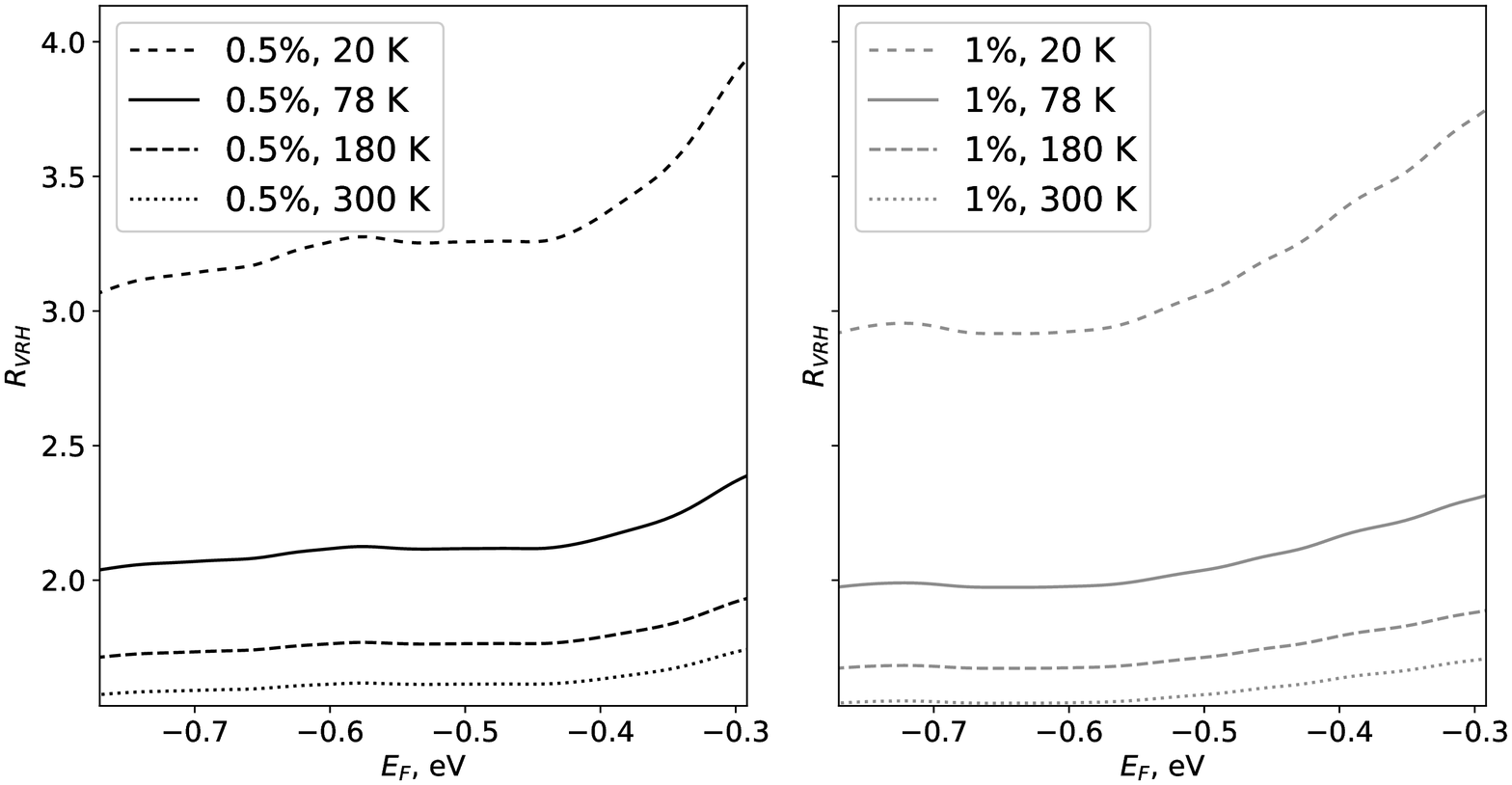} }
\end{center}
\caption{The variable range hopping resistance vs energy in the whole interval (top) and near the peaks of DoS (bottom) for 0.5\% (left column) and 1\% (right column) fluorination content.}
\label{fig7}
\end{figure*}

For fluorinated graphene with relatively high fluorine content, main conductance mechanism is expected to be the hopping conductivity~\cite{fgtuning,assym-fg} due to the localization effects, which is also true for other types of adatoms~\cite{rgo}. The resistance for the case of the hopping conductivity is described with the following empirical relationship:
\begin{equation}
R_{VRH}=R_0 \exp\left[ (\frac{\varepsilon_0}{k_B T})^\alpha \right],\;\varepsilon_0 = \frac{13.8}{\chi^2 \rho(E_F)},\label{Rvrh}
\end{equation}
where $\alpha=1/3$ for the case of 2D variable range hopping (VRH), $k_B$ is the Boltzmann constant,  $\chi$ is the localization length, and $\rho(E_F)$ is the density of states (per eV per area) at the Fermi energy. Taking into account the density of states and neglecting the dependence of the localization length on the temperature and energy, one can try to apply eq.~(\ref{Rvrh}) to estimate the resistivity for the VRH transport mechanism. Fig. \ref{fig7} shows the 2D-VRH resistivity as a function of the gate energy for the constant localization length $\chi\approx$ 10 nm. One can see that the peak in the density of states (see Fig.\ref{fig1}) corresponds to the local minima in the resistivity and vice versa. We also assume that near the peak energy the DoS as a function of energy changes faster than the localization length making the $\chi=const$ approximation valid for this energy region. As for the temperature dependence, one can see that the position of the minima in the resistance as a function of energy (or gap voltage) is fixed for different temperatures, unlike the position of the local maxima. Increase in the fluiorine concentration leads to the more pronounced minima and maxima, appearing at the lower absolute value of energies (or gate voltages), while the increasing temperature leads to the smoothening of falls and peaks, thus leading to the more symmetrical behaviour of the resistance vs the gate voltage. It should be noted that these results are in general agreement to those obtained in the experiment \cite{assym-fg} for the higher concentrations of fluorine at low temperatures ("10 min" and "30 min" fluorination time and 78 K). Nonetheless, the measurements for the low fluorine concentration ("0.5 min") shows no sign of peaks or asymmetry. One can also expect the diffuse transport mechanism to play significant role at very low temperatures, where the Mott-type VRH would be suppressed. The role of the diffuse transport is expected to be significant for electrons, while for the holes the resonant-like scattering of carriers would make it negligible. Thus, we would like to note the different effects of  diffuse and 2D-VRH conductivity mechanisms on the asymmetry of the resistivity: the 2D-VRH mechanism will reduce the hole-type resistance with the energy corresponding to the peak of DOS while the diffuse mechanism will instead decrease the overall electron resistance, and increase the intensity of maxima and minima for the holes. 

\section{Conclusion}
We have numerically calculated the electronic and transport properties of fluorinated graphene with 0.5 and 1~\% fluoride content. We have included the low-lying orbitals for carbon and fluorine atoms into the tight-binding model, that also takes into account the spin-orbit interaction.  We have found the non-resonant peak in the density of states to appear at the energies below the Fermi energy, which is in the general agreement with the analytical   Löwdin-Schrieffer-Wolff model. The position of the peak shifts to the lower energies with the increase of fluorine concentration. We have calculated electrical conductivity and mobility limited by the adatom scattering. For the electrical conductivity, we have found it to have asymmetric behaviour. For the electron conductivity, it has a distinct diffusive behaviour. For the holes, the conductivity is significantly suppressed and is deviating from the standard diffusive behaviour, which can be attributed to the quasi-resonant scattering on the adatoms. A similar pattern can be observed for the carrier mobility as a function of carrier density, where the hole mobility is decreased compared to the electron mobility and the $\mu\sim n^{-1/2}$ behaviour is replaced with more complex non-linear behaviour. 

We have also investigated a different transport regime (the variable-range hopping transport at finite temperatures), and estimated the resistivity as a function of temperature and energy (or gate voltage). Several experiments on the electron transport in fluorinated graphene were performed~\cite{fgtuning,assym-fg}, and in the later case the hole resistance was found to show a non-linear dependence on the gate voltage in some cases. We propose two mechanisms to describe this behaviour. The first one is based on the empiric 2D-VRH law and the presence of the peaks in the density of states for certain values of energies. This leads to a significant decrease of the VRH resistance, which is expected for the corresponding gate voltages. The decrease is more pronounced for lower temperatures, and the value of the gate voltage of the local minima is expected to be temperature-independent. The second mechanism  additionally takes into account the diffuse resistance that leads to the better pronounced local minima and maxima of resistance, as well as to the overall decrease of the resistance for the electrons. While certain features corresponding to the first mechanism were observed in \cite{assym-fg}, manifestations of the second mechanism have not yet been found. One should expect them to be more pronounceable for the lower temperatures and medium fluorine concentrations.

\end{document}